\documentclass{ws-procs9x6}

\begin{document}

\title{Understanding the relationship between the environment of the
black hole and the radio jet: optical spectroscopy of compact AGN}

\author{T.~G. Arshakian\footnote{\uppercase{O}n leave from
\uppercase{B}yurakan \uppercase{A}strophysical
\uppercase{O}bservatory, \uppercase{B}yurakan 378433,
\uppercase{A}rmenia} \footnote{\uppercase{TGA} is grateful to the
\uppercase{A}lexander von \uppercase{H}umboldt \uppercase{F}oundation
for the award of a \uppercase{H}umboldt
\uppercase{P}ost-\uppercase{D}octoral
\uppercase{F}ellowship. \uppercase{T}his work partially supported from
the \uppercase{CONAC}y\uppercase{T} research grants
39560-\uppercase{F} (\uppercase{M}\'exico).} , E. Ros and J.~A. Zensus}
\address{MPIfR, Bonn, Germany\\ E-mail: tigar@mpifr-bonn.mpg.de}
%\address{Max-Planck-Institut f\"ur Radioastronomie,\\Auf dem H\"ugel
%69, 53121 Bonn, Germany\\ E-mail: tigar@mpifr-bonn.mpg.de}

\author{V.~H. Chavushyan} \address{INAOE, Puebla, M\'exico\\ E-mail:
vahram@inaoep.mx} 
%\address{Instituto Nacional de Astrof\'isica Optica
%y Electr\'onica,\\AP 51 y 216, CP 72000, Puebla, Pue., M\'exico\\
%E-mail: vahram@inaoep.mx}

%%%%%%%%%%%%%%%%%%%%%%%%%%%%%%%%%%%%%%%%%%%%%%%%%%%%%%%%%%%%%%
% You may repeat \author \address as often as necessary      %
%%%%%%%%%%%%%%%%%%%%%%%%%%%%%%%%%%%%%%%%%%%%%%%%%%%%%%%%%%%%%%

\maketitle

\abstracts{ We aim to investigate the relationship between radio jet
activity on parsec-scales and the characteristics of both the bright
active galactic nuclei (AGN) and their broad line regions (BLR). For
this purpose, we combine 2cm VLBA$^{\rm a}$ observations of AGN with their
optical spectral observations. This would enable us to investigate the
optical spectra of a set of 172 relativistically beamed, flat-spectrum
AGN with the nuclear disk oriented near to the plane of sky.  Here, we
present first results from optical spectroscopic observations of the
brightest AGN from the 2 cm VLBA survey, and show
a diversity of their spectral morphologies.}

\paragraph{The sample and motivations.}
We intend to combine high-frequency observations of AGN with their
optical spectral observations to study interconnections between the
parsec-scale radio jet properties, central black holes and their
optical environments. For this purpose, we use the sample of compact
radio sources observed at 15 GHz (2 cm) with the VLBA\footnote{Very
Long Baseline Array}. Over 170 sources have been observed since 1994
(see [1,2] for selection criteria and other details).
All AGN are radio loud and core-dominated.  Most of
them possess one-sided jets and superluminal motions on parsec scales.
This can be explained if the jet/counterjet are intrinsically
symmetric and relativistic: the relativistic Doppler boosting favors
the source detection, appearing those as one-sided jet [3,4], and the
small angle between the jet direction and the line of sight leads to
superluminal motions.

Most of the jet viewing angles are small, with a maximum viewing angle
$\sim30^{\circ}$ for quasars and BL Lacs [5]. It implies that most of
relativistically beamed, flat-spectrum AGN have nuclear disks seen
nearly face-on. Therefore, the combination of 2 cm VLBA observations
and optical spectroscopic observations is important for investigating
the spectral properties of AGN, which are not biased by their
orientation.

Our main interests are: (i) to carry out an homogeneous detailed
 spectral classification of AGN and relate it to their radio
 spectral/morphology classification, (ii) to investigate whether the
 properties of the VLBA jets relate to the black hole masses, and
 (iii) to investigate how the jet intrinsic properties relate to the
 geometry/kinematics of BLRs.

\paragraph{Spectral observations.}
We carried out an homogeneous spectral classification of $\sim 70$ AGN
from the 2 cm survey, using an intermediate resolution spectroscopy of
optically bright (m $<$ 17.5) AGN on 2m class GHAO (Cananea, Sonora) and
OAN SPM (Baja California) telescopes. The wavelength coverage
$\sim$3800\AA -- 8000\AA, and spectral resolution 12-15\AA\, allowed us 
to detect a wide range of emission lines going from H$_{\beta}$ to CIV
depending on the redshift of the source. Spectral classification
showed diversity of AGN morphologies: LINERs, Seyfert galaxies, BL
Lacs, quasars and radio galaxies.
% (see selected spectra in the next section).
Based on the NASA Extragalactic Database (NED), within this sample
most of the objects have not a unique and unambiguous
classification. Notice that even using a bright sample of AGN it is
common to find a significative percentage of objects which have been
spectroscopically misidentified [6]. For example, the NED gives
different spectral classifications for 0055+300 (elliptical galaxy,
LINER, Sy3b, and Sy1), while according to our classification it is a
red elliptical galaxy. 

The detailed results will be published elsewhere.

%\section*{Acknowledgments}
%TGA is grateful to the Alexander von Humboldt Foundation for the award
%of a Humboldt Post-Doctoral Fellowship. This work partially supported
%from the CONACyT research grants 39560-F (M\'exico).

%%%%%%%%%%%%%%%%%%%%%%%%%%%%%%%%%%%%%%%%%%%%%%%%%%%%%%%%%%%%%%%%%%%%%%%
% 
%Use this if your figures are put in a subdirectory having the same
%name as the main latex file, ie: 
%
%      ws-procs9x6/procs-fig1.eps      
%      ws-procs9x6/procs-fig2.eps      
%      ws-procs9x6/procs-fig3.eps      
%      ws-procs9x6/procs-fig4.eps      
%      etc.
%
%\begin{figure}[htbp] %ORIGINAL SIZE: width=1.4TRUEIN; height=1.5TRUEIN
%\figurebox{}{}{procf1} %100 percent
%\caption{Labeled tree {\it T}.}
%\end{figure}
%
%%%%%%%%%%%%%%%%%%%%%%%%%%%%%%%%%%%%%%%%%%%%%%%%%%%%%%%%%%%%%%%%%%%%%%%

%\begin{figure}[ht]
%\epsfxsize=5cm   %width of figure - will enlarge/reduce the figures
%\epsfbox{arshakian1-fig1.eps}
%%\figurebox{2cm}{3cm}{} %to have a box alone 
%%\centerline{\epsfxsize=3.7in\epsfbox{arshakian1-fig1.eps}}   
%\caption{. \label{inter}}
%\end{figure}

\end{document}